\documentclass[preprint,showpacs,amsmath,amssymb,showkeys,
floatfix,prd]{revtex4}
\usepackage{amsfonts,amsmath}
\usepackage{graphicx}
\usepackage{dcolumn}% Align table columns on decimal point
\usepackage{bm}% bold math
\usepackage{url}

\DeclareFontFamily{OT1}{pzc}{}
\DeclareFontShape{OT1}{pzc}{m}{it}{<-> s * [1.10] pzcmi7t}{}
\DeclareMathAlphabet{\mathpzc}{OT1}{pzc}{m}{it}

\newcommand{\qslash}{\not{\hbox{\kern-2.3pt $q$}}}
\newcommand{\kslash}{\not{\hbox{\kern-2.3pt $k$}}}
\newcommand{\pslash}{\not{\hbox{\kern-2.3pt $p$}}}

\newcommand{\Pslash}{\not{\hbox{\kern-2.3pt $P$}}}
\newcommand{\Pslashsup}{^\not{\hbox{\kern-0.5pt $^P$}}}

\begin{document}
\title{Astrophysical constraints on millicharged atomic dark matter}

\author {Audrey~K.~Kvam}
 
\author{David C.~Latimer}

\affiliation{Department of Physics, University of Puget Sound,
Tacoma, WA 98416-1031 
}

\begin{abstract}

Some models of inelastic dark matter posit the existence of bound states under some new $U(1)'$ gauge symmetry.  If this new dark photon kinetically mixes with the standard model photon, then the constituent particles in these bound states can acquire a fractional electric charge.  This electric charge  renders a dark-matter medium dispersive.  We compute this frequency-dependent index of refraction for such a medium and use the frequency-dependent arrival time  of light from astrophysical sources to constrain the properties of dark atoms in the medium.  Using optical-wavelength observations from the Crab Pulsar,  we find the electric millicharge of dark (electrons) protons to be smaller than the electric charge $e$ for dark atom masses below 100 keV, assuming a dark fine structure constant $\boldsymbol{\alpha}=1$.  We estimate that future broadband observations of gamma-ray bursts can produce constraints on the millicharge of dark atoms with masses in the keV range that are competitive with existing collider constraints.
\end{abstract}

\maketitle

\section{Introduction}

A concordance of observations support a universe of dark matter (DM), but all concrete evidence for its existence is solely based upon its gravitational interactions \cite{concord1,concord2,wmap9}.  Direct detection experiments aim to rectify this by observing collisions between dark matter and target nuclei.  A host of experiments report no observation of such events, continuing to place more stringent limits on the interaction cross-section.  Yet, the DAMA/LIBRA experiment \cite{dama_libra} reports a statistically significant annual modulation in its detector which could be attributed to the relative motion of the detector through the galaxy's dark matter halo. Furthermore, the CoGeNT  experiment \cite{cogent1} reports  signals above background in its detector which, if due to dark matter, would be consistent with the apparent signal from DAMA/LIBRA.  If these results are due to DM interactions, they occupy
a region of parameter space that has been seemingly ruled out by the CDMS-II \cite{cdmsii_ge} and XENON100 \cite{xe100} experiments.   To reconcile results from DAMA with null results from other experiments, the notion that DM could interact through inelastic channels has been proposed \cite{idm_tuckersmith,idm_chang,idm_marchrussell}.  By fine tuning the inelastic interactions, these models can simultaneously accommodate seemingly conflicting experimental results.

A natural way to incorporate inelastic interactions into a model is to allow DM to be composite, rather than point-like \cite{feng_hiddendm,kaplan_da1,cidm_alves,quirky,formfactor_dm,kaplan_da2,cline_da,kouvaris,wallemacq,cline_strong}.  To create composite particles in a dark sector, modelers introduce a new dark gauge group which results in a binding force among the composite's constituents.  The simplest gauge group is $U(1)$ \cite{feng_hiddendm,kaplan_da1,formfactor_dm,kaplan_da2,cline_da,wallemacq,cline_strong}.  
If the symmetry is unbroken, then the dark photon is massless and can kinetically mix with the Standard Model (SM) photon.  In such models, the particles can effectively couple to the SM photon thereby acquiring a fractional electric charge  \cite{holdom}.  This permits the existence of dark atoms which are overall electrically neutral, but made of constituents with electric millicharge \cite{cline_da,wallemacq,cline_strong}.  

The dark atoms form bound hydrogenic states under the exchange of dark photons, yet their tiny coupling to SM photons opens the possibility that these dark atoms can be probed with SM electromagnetism.  We will explore this possibility in what follows.  As the dark atoms are electrically neutral, they are essentially invisible to low energy photons.  However, for photons which are near the threshold energy needed to transition the dark atom to an excited state, an electromagnetic wave can resolve the charge distribution within the atom.  Considering dark matter as a bulk medium, the interaction between dark matter and light can be characterized in terms of an electric susceptibility and index of refraction \cite{loudon}.   For photon energies below the transition energy $\omega \ll \omega_0$, the medium will rather generically exhibit dispersion quadratic in the photon energy $n(\omega) \approx 1 + A + B\omega^2$ for constants $A$ and $B$.  
Indeed, this is the expected form for the refractive index  from the classical Lorentz model for dielectrics \cite{lorentz}.  Furthermore, this form extends to the realm of particle physics. Combining Fermi's generalized index of refraction formula \cite{fermi} with the low-energy theorems governing the forward scattering amplitude of photons on point-like particles \cite{ggt,low,gg,g} yields the same structure for the refractive index. 

The refractive index for various particulate DM models was considered previously in Ref.~\cite{dm_n}.  In that work, the neutral DM candidate coupled to massive charged particles which interacted with photons.  The process occurred at the one-loop level in perturbation theory and was, thus, highly suppressed resulting in an immeasurable index.  In the case of millicharged dark atoms, the photons can couple directly to the constituents of the DM candidate with measurable consequence.  The dispersive term in the index of refraction can be constrained by astrophysical observation.  If a broadband pulse of light is emitted over a short timescale from an astrophysical object such as the Crab Pulsar or a gamma-ray burst \cite{schaefer}, then a dispersive medium of dark atoms creates a time delay between higher and lower frequency photons.  Limits on a measured time delay can be used to constrain the dispersive term and, thus, the parameters that enter into models of atomic dark matter.

\section{Refractive index of dark atoms}

The atomic dark matter models in Refs.~\cite{cline_da,wallemacq,cline_strong} share a common dark and SM electromagnetic sector.  Two fermions $\psi_\mathbf{p}$ and $\psi_\mathbf{e}$ are charged under the unbroken gauge group $U(1)'$, coupling to the dark photon $\boldsymbol{\gamma}$ with opposite charges $\pm \mathbf{e}$.  [NB: The boldface type is meant to refer to the particles and couplings in the dark sector.]  Kinetic mixing between the dark and SM photons gives rise to an electric millicharge $\pm \epsilon e$ of the dark particles.  The dark ``proton" and ``electron" can form bound states under the dark Coulombic potential $V(r) =  -\boldsymbol{\alpha}/r$, where we define the dark fine structure constant $\boldsymbol{\alpha} := \mathbf{e}^2/ (4\pi)$, and the relative separation between the particles is $r$.   We can use non-relativistic quantum mechanics to describe this dark ``hydrogen", $\mathbf{H}$.  Without loss of generality, we assume $m_\mathbf{e}\le m_\mathbf{p}$ and define the reduced mass $m:= m_\mathbf{e} m_\mathbf{p}/(m_\mathbf{e}+m_\mathbf{p})$.  The relative momentum of the system is  $\mathbf{p} := m\dot{\mathbf{r}}=(m_\mathbf{p} \mathbf{p}_\mathbf{e}-m_\mathbf{e} \mathbf{p}_\mathbf{p})/(m_\mathbf{e} +m_\mathbf{p})$.    From these, we construct the Hamiltonian $H_0 := - \nabla^2/(2m) + V(r)$, which yields the usual  hydrogenic eigenstates $\psi_{n,\ell}$ and energy spectrum $E_n = - \boldsymbol{\alpha}^2 m/(2n^2)$ for quantum numbers $n =1,2,\dots,$ and $|\ell|\le n$. For large values of the dark fine structure constant $\boldsymbol{\alpha}\sim \mathcal{O}(1)$, the fine structure will  introduce significant corrections to the energy levels, and hyperfine structure will become relevant if $m_\mathbf{e}\sim m_\mathbf{p}$.  We will not consider these corrections in what follows.

Classically, a linear dielectric medium, such as  a dilute gas of dark atoms, will acquire a polarization when subjected to an external electric field.  The degree of polarization, or dipole moment per unit volume, can be characterized through the electric susceptibility $\mathbf{P} = \chi_e \mathbf{E}$.  From the susceptibility, we can compute the medium's index of refraction $n =\sqrt{1+\chi_e}$.  To compute the susceptibility of a dark-atom medium, we follow the semiclassical treatment in Ref.~\cite{loudon} in which a classical electromagnetic wave of frequency $\omega$ interacts with the quantum mechanical electric dipole moment of the dark atom $\pmb{\mathpzc{p}} := -\epsilon e \mathbf{r}$.  In the long wavelength limit, $\omega \ll \boldsymbol{\alpha} m =: \mathbf{a}_0^{-1}$, the spatial variation of the electric field is irrelevant, and we incorporate interactions between the wave and the dark atom by including in the Hamiltonian the time-dependent perturbation $H' =-\pmb{\mathpzc{p}} \cdot \mathbf{E}_0 \cos \omega t$.  
 
The electromagnetic wave can induce transitions  between dark atomic energy states.  We argue that the bulk of the dark atoms exist in the ground state.  There are three main mechanisms by which the atoms can be excited beyond the ground state: dark atom self-interactions, absorption of dark photons,  or absorption of SM photons.    The existence of elliptical DM halos severely constrains the DM self-interaction cross section; from the limits in Ref.~\cite{MiraldaEscude:2000qt}, we find that the ratio of the DM self interaction to the dark atom's mass must be $\sigma/m_\mathbf{H} < 0.02$ cm$^2$/g.   These limits can be satisfied
either through tuning the model parameters or dilution of the dark atom component of DM.   For models which satisfy this constraint, we can estimate the mean free time between collisions for dark atoms in the Milky Way's galactic halo.  The mean free path can be estimated as $\lambda \sim (\sigma N)^{-1}$ where $N$ the number density of dark atoms; the number density is related to the DM mass density via $N = \rho/m_\mathbf{H}$.  Then the time between collisions is $t_\text{fp} \sim \lambda/v =  m_\mathbf{H} /(\sigma \rho v) $.  Taking as typical parameters the local dark matter density $\rho \sim 0.3$ GeV/cm$^3$ and $v\sim 200$ km/s, we find $t_\text{fp} \gtrsim 5 \times 10^{18}$ s; i.e., they are non-interacting. 
Dark-atom absorption of dark photons also produce excited states.  The greatest energy density of dark radiation is found in the dark analog of the cosmic microwave background (CMB).  Viable models require the dark radiation to be slightly cooler than the CMB  \cite{adm_cosmo}, so these dark photons will not have sufficient energy to excite the dark atoms.  All that remains is the interaction with SM electromagnetic waves, which we discuss below.

We consider a dark atom in its ground state which can be excited by SM photons.  We  restrict our study to a two-state system, limiting the electromagnetic wave frequency to $\omega \lesssim \omega_0 := E_2-E_1$.  In the presence of the electromagnetic wave, the Hamiltonian for the dark atom is $H= H_0 + H'$, and a general state is $\Psi(t) = c_1(t) e^{-i E_1t} \psi_1+c_2(t) e^{-i E_2t} \psi_2$, with stationary eigenstates $\psi_{1,2}$.  Using the Schr\"odinger equation, we can develop coupled differential equations for the coefficients $c_1$ and $c_2$.  These equations must be amended to account for spontaneous emission of a dark photon from the excited state--a field theoretic result.  With this extra term proportional to the decay constant $\Gamma$, we have the equation for $c_2$
\begin{equation}
\frac{\mathrm{d}}{\mathrm{d}t} c_2(t) = -i \, (\pmb{\mathpzc{p}}_{21}\cdot \mathbf{E}_0) e^{i\omega_0t} \cos(\omega t) c_1 (t)   -\Gamma c_2(t)
\end{equation}
where $ \pmb{\mathpzc{p}}_{21} := \langle \psi_2 | \pmb{\mathpzc{p}}  | \psi_1  \rangle$ and $\Gamma=  \frac{2^7}{3^8}\boldsymbol{\alpha}^5 m$, adapted from atomic hydrogen \cite{loudon}.

We solve perturbatively for the coefficients $c_{j}(t) = c_{j}^{(0)}(t)+ c_{j}^{(1)}(t)+\cdots$  with initial conditions $c_1(0)=1$ and $c_2(0)=0$.  To determine the {\em linear} response of the atom to the field $E_0$, we only need the zeroth order approximation for $c_1 \approx 1$ and the first order approximation for $c_2$
\begin{equation}
c_2(t) \approx -\frac{1}{2} (\pmb{\mathpzc{p}}_{21}\cdot \mathbf{E}_0) \left[ \frac{e^{i(\omega_0 + \omega)t}}{\omega_0 + \omega - i \Gamma}+\frac{e^{i(\omega_0 - \omega)t}}{\omega_0 - \omega - i \Gamma }\right].
\end{equation}
The induced dipole moment for $\Psi$ is thus
\begin{equation}
\pmb{\mathpzc{p}}(t) = -\epsilon e \langle \Psi(t) | \mathbf{r} | \Psi(t) \rangle = -2\epsilon e\,  \mathrm{Re} [ \mathbf{r}_{12} c_1^*c_2 e^{-i \omega_0 t}].
\end{equation}
We note that the polarization $P(t)$ will be the product of the average induced dipole moment in the direction of $\mathbf{E}_0$ and the number density $N$.  Averaging over the relative orientation between $\mathbf{r}_{12}$ and $\mathbf{E}_0$,   the susceptibility is
\begin{equation}
\chi_e = N\pi\frac{2^{18} }{3^{11}}  \frac{\epsilon^2 \alpha }{\boldsymbol{\alpha}^2 m^2}\frac{\omega_0}{\omega_0^2 -(\omega+i\Gamma)^2}.  
\end{equation}

Given that the DM medium is weakly interacting and dilute, $\chi_e$ is nearly zero.  As such we can approximate the index of refraction as $n \approx 1 +\frac{1}{2} \chi_e$.  For frequencies below the transition energy $\omega < \omega_0$, the index of refraction exhibits normal dispersion; that is, $n(\omega)$ increases with $\omega$.  In fact, for $\omega \ll \omega_0$, the dispersion is quadratic in frequency
\begin{equation}
\mathrm{Re}(n) \approx 1  +  N\pi\frac{2^{20} }{3^{12}}  \frac{\epsilon^2 \alpha }{\boldsymbol{\alpha}^4 m^3}\left( 1 + \frac{\omega^2}{\omega_0^2}\right) ,
\end{equation}
neglecting the small term proportional to $\Gamma^2$.

\section{Limits from the Crab Pulsar}

As a broadband pulse of electromagnetic radiation travels through a dispersive medium, the pulse shape spatially broadens since the phase speed of each component wave is frequency dependent.  For a medium which is normally dispersive, higher frequency components of the pulse will lag the lower frequency components.  Over large distances, a time lag can accrue between these two components.  For a dispersive medium of atomic DM, the time lag between simultaneously emitted waves of frequency $\omega_\text{hi}$ and $\omega_\text{lo}$ traveling over a baseline $\ell$  is
\begin{equation}
\tau \approx \ell N\pi\frac{2^{20} }{3^{12}}  \frac{\epsilon^2 \alpha }{\boldsymbol{\alpha}^4 m^3}\left( \frac{\omega_\text{hi}^2 -\omega_\text{lo}^2}{\omega_0^2}\right), \label{time_lag}
\end{equation}
assuming $\omega_\text{hi}\ll \omega_0$.

\begin{figure}[t]
\includegraphics[width=8.6cm]{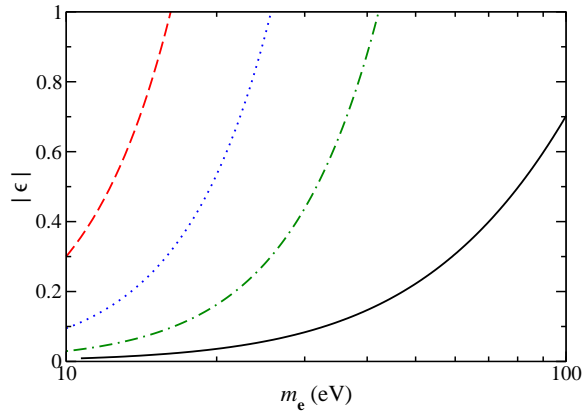}
\caption{(Color online) Constraint on the electric millicharge $|\epsilon| e$ of the dark electron and proton from the Crab Pulsar dispersion limits with $\boldsymbol{\alpha}=1$.  The dashed (red) curve has $m_\mathbf{p}=100$ keV; the dotted (blue) curve has $m_\mathbf{p}=10$ keV; the dot-dashed (green) curve has $m_\mathbf{p}=1$ keV; and the solid (black) curve has $m_\mathbf{p}=100$ eV.  The parameter space above each curve is excluded by the data.\label{fig1}}
\end{figure}

The frequency dependence of the speed of light has been severely constrained in the optical spectrum from the observation of radiation coming from the Crab Pulsar.  The phase difference between 0.35 and 0.55 $\mu$m radiation was determined to be less than 10 $\mu$s \cite{crab_pulsar}.  Since the pulsar is 2 kpc from the Earth, this provides a stringent limit on the ratio $\tau/\ell \approx 5 \times 10^{-17}$.    We will assume that the entirety of dark matter in the galactic halo is atomic DM, and to yield the most conservative constraints on the properties of dark atoms, we will assume that any dispersive effects suffered by light are due exclusively to the dispersion from atomic DM.  Since our limits are only valid for $\omega \ll \omega_0$, the high frequency observation places bounds on the product of the dark fine-structure constant and the system's reduced mass $\boldsymbol{\alpha}^2 m \gg \frac{8}{3} \omega_\text{hi} = 9.5$ eV.  We will assume the local DM mass density to be $\rho = 0.3$ GeV/cm$^3$, and then the DM number density is $N = \rho/m_\mathbf{H}$ where $m_\mathbf{H} = m_\mathbf{p} + m_\mathbf{e} -\frac{1}{2}\boldsymbol{\alpha}^2 m$.

The most severe constraints on the properties of the dark atoms will come from photons whose energies are close to the transition energy, i.e., $\omega_\text{hi} \sim \omega_0$.  In the case of the Crab Pulsar, we have $\omega_\text{hi} =3.5$ eV, which limits the product $\boldsymbol{\alpha}^2 m$ to be on the order of tens of electronvolts.  For $\boldsymbol{\alpha} =1$, the most massive dark atom that we can meaningfully constrain has a dark proton mass $m_\mathbf{p} = 1.0 $ MeV; then for a dark electron mass of 9.5 eV, the electric millicharge of these particles satisfies $\epsilon < 0.84$.  For smaller dark proton masses, there is a broader range of dark electron masses that leads to meaningful constraints on the millicharge, Fig.~\ref{fig1}.  For a dark proton mass of 100 keV, a dark electron mass of $10$ eV provides the most stringent constraint on the millicharge, $|\epsilon| < 0.3$.  Limits improve with lighter atoms.  For example, a dark atom with $m_\mathbf{p} = 100$ eV and $m_\mathbf{e} = 20$ eV results in the constraint $|\epsilon|< 0.04$.  

The dark atomic masses constrained by dispersion are much smaller than those of the dark matter candidates in Ref.~\cite{cline_da}, $\sim 6$ GeV, and Ref.~\cite{wallemacq}, $\sim 650$ GeV.  Indeed, the region of parameter space which we exclude is already ruled out by the self-interaction limits found in Ref.~\cite{MiraldaEscude:2000qt}.
Furthermore, stringent limits on the existence of millicharge particles, cf.~Refs.~\cite{davidson,Vogel:2013raa}, would rule out the dark protons and electrons constrained by our dispersion limits. 

\section{Potential limits from GRBs}

Observations of the Crab pulsar place severe constraints upon the dispersive nature of the interstellar medium in the Milky Way, 
yet these constraints are only relevant for a narrow band of the electromagnetic spectrum.  Even more severe constraints can be derived from distant sources like gamma ray bursts (GRBs), and these constraints span a much broader spectral range.  
GRBs produce an intense burst of high-energy photons on a short time scale which can be viewed over cosmological distances, increasing the effective baseline, $\ell$ in   Eq.~(\ref{time_lag}), between the source and observer.  The high-energy photons from GRBs, upwards of 100 GeV, have the potential to constrain composite millicharged dark matter of greater mass, and the associated optical, infrared, and radio afterglow allow for their use in broadband observations.  The high energy photons from distant GRBs have already been used to place lower bounds on the scale at which Lorentz-invariance violating effects could modify the photon's dispersion relation \cite{AmelinoCamelia:1997gz,ellis,Boggs:2003kxa,Ellis:2005wr,jacobpiran,ellis2,Ellis:2009yx,AmelinoCamelia:2009pg,Vasileiou:2013vra}.  Along similar lines, we will consider how the (non-)observation of dispersion, attributable to dark matter, impacts the properties of composite dark matter.

Relative to the Crab pulsar, using GRBs for dispersion studies has two drawbacks:  the intergalactic DM density is much smaller and the temporal profile of the source is not known.  With regards to the DM density, the Crab pulsar is within the Milky Way where the dark matter halo density is relatively large.  For distant GRBs, the average DM density is $\rho \simeq 1.20\times 10^{-6}$ GeV/cm$^3$\cite{wmap9}, a factor of $10^5$ smaller than our local density.  
In terms of timing resolution, the pulsar is well suited for dispersion studies because its regular rotational period means the temporal profile of light emitted from these objects is well known.  GRBs do not share this feature; the spectra of the bursts  \cite{Gruber:2014iza} and their lower frequency afterglows \cite{Meszaros:1996sv,Cenko:2010cg,Zaninoni:2013hca}  are varied.  Since dispersion measurements rely on temporal knowledge of the emission spectra, a single GRB event cannot yet be used to constrain DM properties.  But with a large number of GRB observations located at a variety of redshifts, it is expected that, statistically, random variations amongst the sources should wash out, and any real effects that indicate dispersion should survive.  The Fermi Gamma-Ray Space Telescope \cite{fermi_lat} is  dedicated to the task of gamma-ray observations, so that in the future, a sufficient number of GRBs may be used to provide meaningful constraints.  Despite the present dearth of data, we will estimate the possible reach of such a study based upon the observation of a single burst.     

Since GRBs can be viewed at cosmological distances, the frequency-dependent time-lag accrued by simultaneously emitted pulses of light traveling through a dispersive medium of dark atoms requires modification due to the universe's expansion \cite{jacobpiran}.  First, the light-travel time between source and detector is dependent upon the redshift of the source and the local expansion rate.  Second, at redshift $z$ the number density of dark matter increases by a factor of $(1+z)^3$.  Finally, as we look into the past, the wavelength of light blue shifts relative to its value $\omega$ at the detector (at $z=0$). Incorporating these three factors, the time lag accrued between (detected) frequencies $\omega_\text{hi}$ and $\omega_\text{lo}$ simultaneously emitted from a source at redshift $z$ becomes
\begin{equation}
\tau \approx N\pi\frac{2^{20} }{3^{12}}  \frac{\epsilon^2 \alpha }{\boldsymbol{\alpha}^4 m^3}\left( \frac{\omega_\text{hi}^2 -\omega_\text{lo}^2}{\omega_0^2}\right) \int_0^z \frac{(1+z')^5 \mathrm{d}z'}{H(z')}   
\end{equation}
where $N$ is the present-day DM number density and  the Hubble expansion rate at redshift $z'$ is $H(z') = H_0 \sqrt{(1+z')^3\Omega_M + \Omega_\Lambda}$. 
The cosmological parameters determined through the combined analysis of WMAP nine-year data 
in the $\Lambda$CDM model with  distance measurements from Type Ia supernovae (SNe) and with baryon acoustic oscillation
information from the distribution of galaxies can be found in Ref.~\cite{wmap9}. The Hubble constant
today is $H_0 = 69.32 \pm 0.80 \,\hbox{km/s/Mpc}$, whereas 
the fraction of the energy density
in matter relative to the critical density today is $\Omega_M = 0.2865^{+0.0097}_{-0.0096}$ and 
the corresponding fraction of the energy density 
in the cosmological constant $\Lambda$ is 
$\Omega_\Lambda = 0.7135^{+0.0095}_{-0.0096}$~\cite{wmap9}. 

The maximum reach for idealized GRB constraints are derived from observations of the most distant events. There have been two GRB observations with confirmed redshift $z>8$:  GRB 090423 with $z=8.3\,$ \cite{Tanvir:2009zz} and GRB 090429B with $z=9.4$ \cite{Cucchiara:2011pj}.  We will use as our exemplar the observation of GRB 090423 because, in addition to the burst detection, the afterglow was observed down through the radio band \cite{grb090423_radio}.  The strength of the pulsar constraint relied on the fact that the detector's temporal resolution was $\sim 10 \mu$s.  This same resolution is possible for the Fermi Large Area Telescope which detects gamma rays from 20 MeV up to 300 GeV \cite{fermi_lat}; however, this resolution is likely not possible for the variety of ground based telescopes which observe the GRB afterglows.  For example, the integration times for the ground-based Gamma-Ray Burst Optical and Near-Infrared Detector are on the order of tens to hundreds of seconds \cite{grond}.  Longer integration time weakens the ability to constrain dispersion for large values of the dark fine-structure constant $\boldsymbol{\alpha}$, but meaningful constraints are realizable in the radio band. We estimate such constraints for GRB 090423 for a small value of $\boldsymbol{\alpha}$.

The most competitive limits arise whenever the observed high-frequency photons are near the threshold energy $\omega_\text{hi} \sim\omega_0$; however, we must be mindful that the wavelength blueshifts as we look into the past.  In order to not surpass the threshold energy at redshift $z$, we must limit the measured high frequency photons to $\omega_\text{hi} \sim (1+z)^{-1} \omega_0$.  Assuming this, along with a large separation between high and low frequency bands $\omega_\text{hi} \ll \omega_\text{lo}$, we find limits on the electric millicharge of the dark electron to be
\begin{equation}
\epsilon^2  \lesssim \frac{(1+z)^2\tau}{2\pi \alpha \rho }  {\boldsymbol{\alpha}^4m_\mathbf{H} m^3} \left[ \int_0^z \frac{(1+z')^5 \mathrm{d}z'}{H(z')}\right]^{-1}   
\end{equation}
Consulting Ref.~\cite{grb090423_radio}, we find that the location of GRB090423 was observed by several radio telescopes but detection of the afterglow was only confirmed by the Plateau de Bure Interferometer \cite{GCN9273} (at 100 GHz) and the Very Large Array (VLA)\cite{grb090423_radio} (at 8.46 GHz).  These detections were separated by eight days, so they cannot provide a meaningful constraint.  What is needed is near simultaneous multi-band detection of the GRB; at best, we can make a reasonable estimate of the reach of such a detection.  We suppose that a high frequency observation is made by the VLA at 8.46 GHz while a lower frequency observation is made by another telescope.  Typical integration times for these telescopes are on the order of tens of minutes, but for our purposes, we suppose a somewhat optimistic integration time of 100 s.  Then, if the detection of the GRB can be established at 8.46 GHz and a lower frequency within this 100 s window, we find that this measurement could constrain the electric millicharge of a dark atom with masses $m_\mathbf{e} = m_\mathbf{p} = 1$ keV and $\boldsymbol{\alpha}=5\times10^{-4}$ to be $\epsilon \le 2.8\times 10^{-4}$. 

This estimate is competitive with terrestrial constraints on the existence of millicharged particles derived from colliders.  For particle masses below 200 MeV, the SLAC electron beam-dump experiment \cite{Donnelly:1978ty,slac_ebd} limits the particle's millicharge to be $\epsilon \le 3\times 10^{-4}$ \cite{Davidson:1991si}.  On the other hand, millicharged particles with masses in the keV range would disrupt big-bang nucleosynthesis in the early universe; constraints upon the light degrees of freedom would preclude the existence of these millicharged particles \cite{Davidson:1991si}.

\section{Conclusion}

Dark matter models which consist of composite states under some unbroken $U(1)'$ gauge symmetry can explain possible DM signals in the DAMA/LIBRA and CoGeNT experiments that are seemingly inconsistent with the null results from the CDMS-II and XENON100 experiments.   In some models, the dark photon mixes with the standard model photon resulting in electric millicharges for the particles in the dark sector.    This interaction with the SM photon results in non-trivial optical properties for a dark-matter medium.  We computed the electric susceptibility of a medium of dark atoms and determined that such a medium is dispersive.  The dispersion can result in time lags between simultaneously emitted pulses of light of differing frequencies.  

Using stringent limits on the (optical) frequency dependence of the speed of light from the Crab Pulsar, we are able to constrain the properties of a dark matter medium consisting of dark atoms.  For a dark fine structure constant $\boldsymbol{\alpha}=1$, we find that we can limit the electric millicharge of dark (electrons) protons to be less than the SM proton charge, $| \epsilon| \le 1$, for dark atom masses below 100 keV.  Millicharged particles in this mass range have been ruled out by collider searches and interact too strongly to be consistent with the dark matter haloes of elliptical galaxies.  GRBs yield more stringent constraints on the dispersion of the cosmos.  Given their brightness and broad spectral range, these objects are promising sources by which we can constrain millicharged atomic dark matter.  We estimate that observations of the radio afterglow could result in constraints on the electric millicharge of the dark electron and proton for masses in the keV range that are competitive with collider constraints; however, big-bang nucleosynthesis would rule out the existence of these light particles in the early universe.  

The technique that we develop in this work, namely that cosmic dispersion can constrain dark atomic  properties, is applicable to other models of composite dark matter that have millicharged constituents.  At present, we know of no other models of composite dark matter with millicharged constituents.  Likely, this is due to parsimony; i.e., the composite system is bound under some broken gauge group so that charged constituents under a new $U(1)'$ gauge group would be superfluous.  However, the dark sector could be as complex as the Standard Model sector, and the constituents of the composite dark matter could indeed couple to dark photons, opening the possibility that they acquire an electric millicharge.  If this is the case, then dispersive constraints would impact the parameters of the composite dark matter model.

\section{ACKNOWLEDGMENTS}

AK acknowledges partial support during the completion of this work from the  Washington NASA Space Grant Consortium and  the University of Puget Sound.  DL thanks Bernie Bates for useful discussions on radio astronomy.

\bibliographystyle{ws-mpla}

\end{document}